\documentclass[aps,prl,reprint,groupedaddress]{revtex4-1}

\usepackage{float}
\usepackage{hyperref}
\usepackage{epsfig}
\usepackage{graphicx}
\usepackage{subfigure}
\usepackage{latexsym}
\usepackage{color}
\usepackage{fullpage}
\usepackage{epstopdf}
\usepackage{dcolumn}
\usepackage{bm}
\usepackage{ulem}
\usepackage{units}
\usepackage{amsmath}
\usepackage{lineno}
\usepackage{verbatim}

\begin{document}

\title{Corrugation in a molecular C$_{60}$ monolayer as a frustrated magnet}

\author{M. Alfonso-Moro, V. Guisset, P. David, B. Canals, J. Coraux, and N. Rougemaille} 

\affiliation{Univ. Grenoble Alpes, CNRS, Grenoble INP, Institut NEEL, 38000 Grenoble, France}

\date{\today}

\begin{abstract}
Under certain experimental conditions, the deposition of C$_{60}$ molecules onto an atomically flat copper surface gives rise to the formation of corrugated islands.
This corrugation, which reflects a molecular displacement perpendicular to the surface plane, presents an astonishing pattern: it is well described by a frustrated Ising spin Hamiltonian whose thermodynamics is compatible with a spin liquid about to transit towards an ordered zigzag state.
Here we study the statistical properties of such a molecular corrugation using tools generally employed in frustrated magnetism.
More specifically, the real and reciprocal space analysis of pairwise molecule correlations allows us to demonstrate that the C$_{60}$/Cu system, in which magnetism is totally absent, has all the characteristics of a triangular Ising antiferromagnet.
Our results indicate that the organization of two-dimensional matter, at the molecular length scale, sometimes turns out to be particularly close to that encountered in highly frustrated magnets.
\end{abstract}

\maketitle

\textit{Introduction}-- 
In condensed matter physics, frustrated magnetism has established itself as an important branch of modern magnetism \cite{BookGingras, BookRamirez}. 
If the concept of frustration \cite{Toulouse1977} can be introduced in different ways, it is commonly associated with the extensive degeneracy of the low-energy physics characterizing certain lattice models \cite{Baxter1982, Lieb2004}, magnetic compounds \cite{BookGingras, BookRamirez, Gardner2010} or artificial structures \cite{Nisoli2013, Rougemaille2019, Heyderman2020}. 
Because of its property to delay or even suppress magnetic ordering \cite{Balents2010}, frustration is at the origin of many unconventional behaviors of matter. 
Interestingly, that was in a non-magnetic system -- water ice in its usual hexagonal form -- that the macroscopic degeneracy of a ground state manifold was first reported \cite{Giauque1936} and interpreted \cite{Bernal1933, Pauling1935}. 
But unlike magnetic systems, which have seen the emergence of a vast variety of compounds with exotic phases and excitations \cite{Castelnovo2008, Kadowaki2009, Jaubert2009, Henley2010, Mengotti2011, Perrin2016, Farhan2019}, very few non-magnetic systems like water ice exhibit a highly degenerate low-energy physics.

Several notable exceptions in surface science suggest that certain non-magnetic surface alloys \cite{Ganz1991, Carpinelli1998, Ottaviano2003}, two-dimensional materials \cite{Azizi2020} or non-magnetic molecular assemblies on surfaces \cite{Charra1998, Pai2004, Gardener2009, Pasens2015} may possibly share common features with the triangular Ising antiferromagnet (TIAF). 
This Ising spin model, intensively studied in the early 50s \cite{Houtappel1950, Husimi1950, Newell1950, Temperley1950, Wannier1950}, is the archetype of a spin liquid: fluctuating down to the lowest temperatures, it is characterized by a zero-point entropy. 
As a matter of fact, the methodology and terminology often used to investigate highly frustrated magnets are rarely employed in surface physics, leaving no clear evidence that the organization of non-magnetic matter gives rise to a spin liquid-like physics.

\begin{figure*}
\includegraphics[width=164 mm]{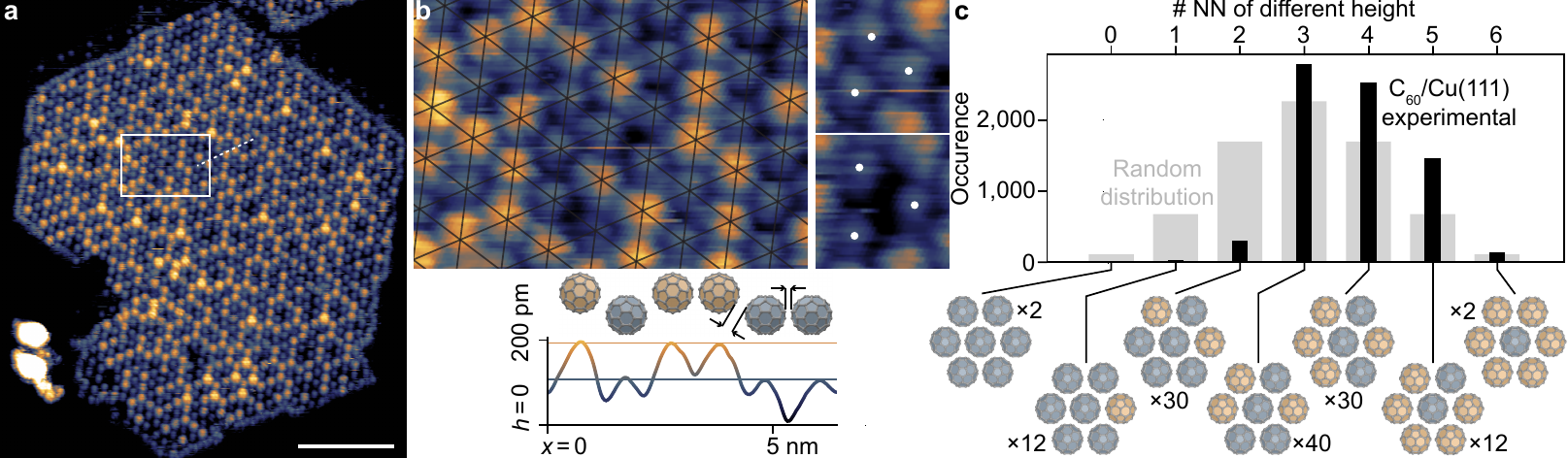}
\caption{\label{fig1} (a) STM image of a typical C$_{60}$ island on a Cu(111) surface. Molecules appear as bright/dark dots depending on their height. The island contains about 1,000 molecules. Scale bar is 10 nm. (b) Zoom-in image showing the closed-pack nature of the molecular arrangement. The molecules sit on the nodes of a triangular lattice (highlighted in black). A slight increase of the lattice parameter is often observed when three low-lying molecules sit next to each other (right panels). A height profile along the dashed line in (a) is shown together with a side-view cartoon of a row of molecules. Two specific molecular heights are observed, separated by about 85 pm. (c) Histogram counting nearest neighbor molecules having different height (in black), averaged over ten STM images (about 6,200 molecules). The histogram expected for a random arrangement is shown in grey. The cartoons illustrate the configuration types associated with the histogram values, taking the central molecule of the 7-molecule plaquette as a reference. All other configurations can be derived by color/arrangement permutations. Numbers indicate the configuration degeneracy.}
\end{figure*}

Here we study a two-dimensional non-magnetic system under the prism of magnetic frustration. 
Combining scanning tunneling microscopy and Monte Carlo simulations, we demonstrate that a monolayer of C$_{60}$ fullerene molecules deposited on a copper surface presents a corrugation whose properties are well captured by a frustrated Ising spin Hamiltonian on a triangular lattice. 
Analyzing pairwise spin correlations in real and reciprocal space, we show that the spin Hamiltonian \textit{must} contain at least three antiferromagnetic coupling terms to account for the experimental findings.
Moreover, Monte Carlo simulations allow us to bound the relative values of these three coupling terms, providing key information on the molecular assembly. 
We also show that the physics we observe is representative of a correlated disorder, at thermodynamic equilibrium, and at an effective temperature well below the interaction strength coupling nearest molecules. 
In other words, our work introduces C$_{60}$ assemblies as a non-magnetic frustrated system, and demonstrates the relevance of applying a methodology inherited from frustrated magnetism in surface science.

\bigskip

\textit{Lattice geometry, Ising degree of freedom and antiferromagnetic couplings}-- 
The system we consider is a submonolayer of C$_{60}$ molecules deposited onto an atomically flat Cu(111) surface.
The molecules are evaporated under ultra-high vacuum on a Cu single crystal kept at room temperature (see Supplemental Material \cite{SuppMat}).
Relatively large islands of C$_{60}$ molecules are subsequently imaged using scanning tunneling microscopy (STM).
A typical STM image of a C$_{60}$ island containing about 1,000 molecules is shown in Fig.~\ref{fig1}a \cite{note_phase}.
Consistent with previous works \cite{Pai2004}, the molecular arrangement is closed-pack and the fullerenes sit on a triangular lattice (see Fig.~\ref{fig1}b where the triangular lattice is highlighted).
Key for this work, two STM contrasts are observed for the C$_{60}$ molecules, which appear either bright or dark.
Although STM measurements always convolute topographic and electronic information, the main origin of the contrast is topographic \cite{Pai2010}, and the bright/dark contrast reveals a height difference.
As shown by the topographic line scan reported in Fig.~\ref{fig1}b, this height difference is $85 \pm 17 $pm.
Strikingly, the molecules bulge out of the surface at two specific height values (bright and dark colors in Figs.~\ref{fig1}a-b), such that an Ising degree of freedom can be assigned to the molecule height \cite{Maria2023, note_contrast}.
From the STM data, we find an inter-molecular in-plane distance of $\sim$ 9.7$\AA$, typically 3$\%$ smaller than the bulk value ($\sim$ 10.0$\AA$).
The C$_{60}$ islands are under compressive strain and buckle \cite{note_buckling}.

\begin{figure*}
\includegraphics[width=130 mm]{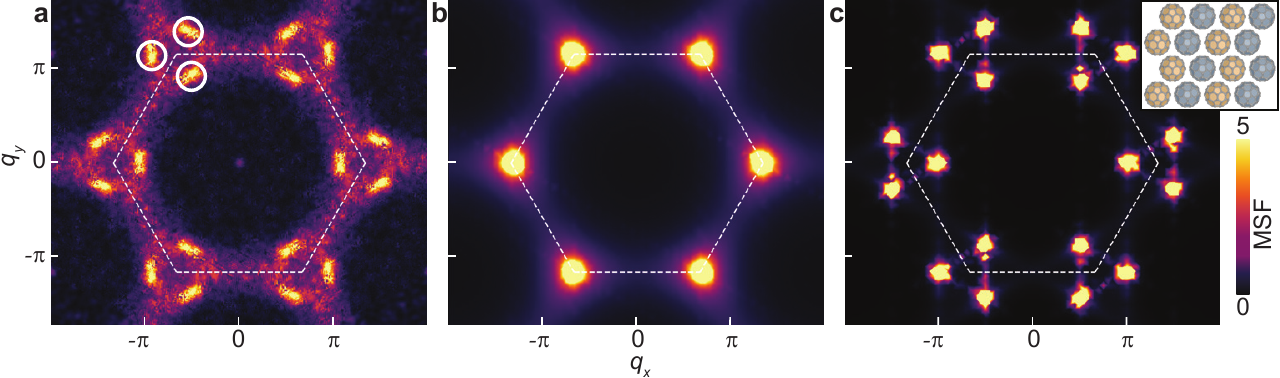}
\caption{\label{fig2} (a) Experimental magnetic structure factor averaged over 10 STM images. The Brillouin zone is highlighted by a white dashed line. (b,c) Numerical magnetic structure factors deduced from Monte Carlo simulations in: the low-temperature regime of the (short range) triangular Ising antiferromagnet and (b) within a zigzag magnetic order state, represented in the inset (c). The intensity scale is the same in all three MSF.}
\end{figure*}

Our STM images also reveal that the lateral distribution of the corrugation is not random, but rather exhibits \textit{correlated} disorder.
C$_{60}$ molecules are generally surrounded by more molecules having a different height than having the same one.
In other words, the Ising degree of freedom describing the C$_{60}$ state is, on average, alternating between neighboring sites.
This property can be quantified by plotting the histogram counting pairwise molecules having distinct height (see Fig.~\ref{fig1}c).
Each molecule having six nearest neighbors, the values range from $0$ to $6$.
A random distribution would lead to a symmetric histogram centered around $3$.
Instead, our data yield a non symmetric histogram centered around a mean value of $3.8$.

Taken as a whole, these observations lead to an important result: the three main ingredients leading to geometrical frustration are present in our system.
An Ising degree of freedom can be assigned to the molecule position (up or down, Fig.~\ref{fig1}a), these molecules are interacting in an antiferromagnetic fashion (Fig.~\ref{fig1}c), and they sit on a triangular lattice (Fig.~\ref{fig1}b).
This C$_{60}$/Cu(111) phase thus has all the ingredients of the triangular Ising antiferromagnet, the canonical example of a frustrated magnet being a spin liquid in its ground state, but here in a non-magnetic system.

The question then arises to what extent the molecular assemblies we imaged are snapshots of arrested spin liquid configurations.
To address this question, in the rest of the paper we describe the molecular assemblies in terms of effective magnetic states, in which the Ising variable is considered to be an ``up'' or ``down'' spin.

\bigskip

\textit{Disorder and magnetic correlations}--
To characterize the corrugation disorder, we used a tool commonly employed to study frustrated magnetism and computed the so-called magnetic structure factor (MSF, see Supplemental Material \cite{SuppMat}) from the real space images \cite{Rodriguez-Carvajal, Rougemaille2021}.
To improve statistics, the MSF is averaged over 10 STM images (for a total of about 6,200 molecules, see Supplemental Material \cite{SuppMat}).
The average MSF is presented in Fig.~\ref{fig2}a.
The MSF has two important characteristics: a diffuse structured pattern, typical of a correlated disorder, and low-intensity regions centered at specific wave-vectors resembling emerging magnetic Bragg peaks (highlighted by white circles in Fig.~\ref{fig2}a).
Contrary to the first impression given by real space (STM) images, the molecular assemblies show some degree of order, at variance with expectations from the (nearest neighbor) TIAF model (see Fig.~\ref{fig2}b).
Although the two MSFs (Figs.~\ref{fig2}a and ~\ref{fig2}b) share a common diffuse background intensity, distinct magnetic correlations clearly develop in our system, especially close to the corners of the first Brillouin zones.

We can then wonder to what ordered pattern these emerging Bragg peaks correspond to.
Consistent with recent theoretical works \cite{Mila2016}, we find that a zigzag magnetic pattern yields Bragg peaks precisely where we observe strong MSF intensities (compare Figs.~\ref{fig2}a,c). 
Although our molecular assemblies seem essentially liquid-like, they start developing zigzag correlations.
It is important to note that a zigzag pattern is obtained in the TIAF model only when interactions beyond second neighbors are involved \cite{Tanaka1976, Kaburagi1978, Takagi1995, Mila2016}.
Hence, if an Ising spin Hamiltonian potentially describes our experimental findings, this Hamiltonian cannot be short-ranged like in the seminal TIAF model introduced by Wannier \cite{Wannier1950}. 
Thus, C$_{60}$ molecules cannot be described by a hard-sphere model that buckles under compressive strain.

\bigskip

\textit{Coupling strengths}--
Including spin-spin couplings up to the third nearest neighbors ($J_{1,2,3}$) in the Ising spin Hamiltonian is sufficient to stabilize a zigzag ground state \cite{Mila2016}. 
Within the ranges of possible coupling values, the following question now arises: which $\{J_1,J_2,J_3\}$ triplets allow describing the properties of the C$_{60}$ layer? 
Below, we show how such triplets can be found, and demonstrate three key results:

\noindent 1) We can define an Ising spin Hamiltonian whose low-energy properties describe very well our experimental findings,  

\noindent 2) The values of the first three coupling strengths of this Hamiltonian can be bounded, 

\noindent 3) The molecular assemblies reflect a physics at thermodynamic equilibrium and an effective temperature can be assigned to the imaged C$_{60}$ configurations.

\noindent To do so, we compared the real space pairwise spin correlations we deduced from our STM images to predictions from a set of Monte Carlo simulations in which the three coupling strengths are varied.
The spin Hamiltonian we consider has the form $H=-J_1 \sum_{\langle i, j \rangle} \sigma_i \sigma_{j}-J_2 \sum_{\langle\langle i, j \rangle\rangle} \sigma_i \sigma_{j}-J_3 \sum_{\langle\langle\langle i, j \rangle\rangle\rangle}\sigma_i \sigma_{j}$, where $\sigma_i$ is an Ising variable on the site $i$ of a triangular lattice, and $J_1, J_2, J_3$ are the first, second and third neighbor coupling strengths, respectively.
To reduce the number of free parameters and without loss of generality, $J_1$ is set to $-1$ ($J_1$ being negative to account for the antiferromagnetic coupling deduced from the measurements, see Fig.~\ref{fig1}c).
We then compute the thermodynamic properties of the spin Hamiltonian for a set of $J_2$ and $J_3$ coupling strengths, and we compare the values of the second ($C_2=\left\langle\sigma_i \sigma_{i+2}\right\rangle$) and third ($C_3=\left\langle\sigma_i \sigma_{i+3}\right\rangle$) pairwise spin correlators in the low-temperature regime to the values we measured (see Supplemental Materials \cite{SuppMat}).
Results are reported in Figs.~\ref{fig3}a,b.

\begin{figure}
\includegraphics[width=80mm]{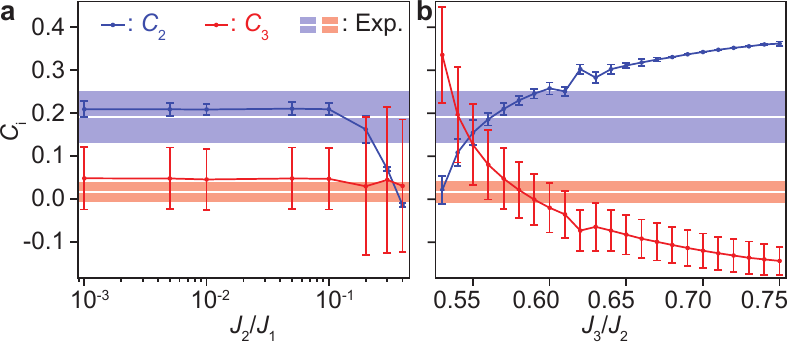}
\caption{\label{fig3} Numerical values of the second ($C_2$, blue) and third ($C_3$, red) spin-spin correlations and their corresponding standard deviations (vertical bars) obtained at low temperature as a fonction of (a) $J_2/J_1$ (with $J_3/J_2$=0.57) and (b) $J_3/J_2$ (with $J_2/J_1$=0.1). The corresponding experimental values are represented by white lines (average) and shaded regions (standard deviation).}
\end{figure}

\begin{figure*}
\includegraphics[width=130mm]{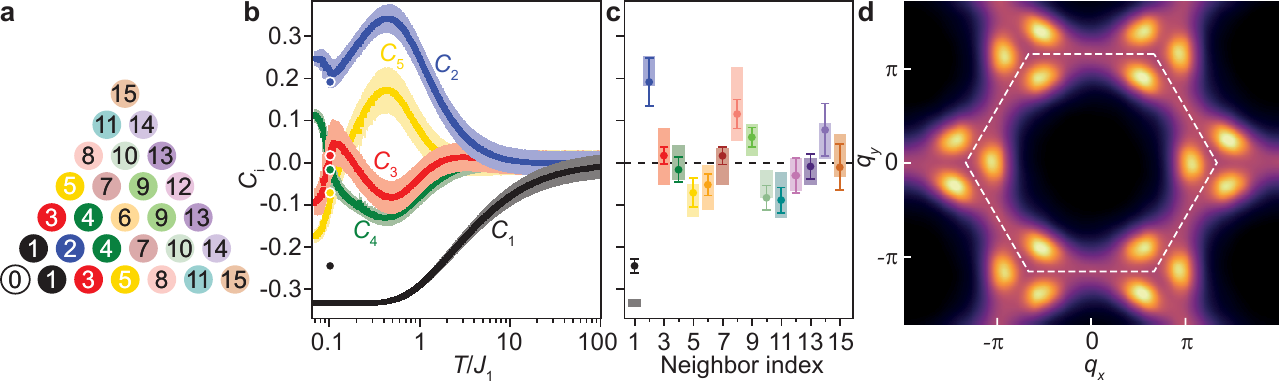}
\caption{\label{fig4} (a) Schematics of the first fifteen neighbors. (b) Temperature dependence of the first five spin-spin correlators and their corresponding standard deviations deduced from Monte Carlo simulations. Experimental values are represented by colored dots. (c) Experimental (dots) and numerical values (colored bars) of the first fifteen spin-spin correlators. (d) Numerical magnetic structure factor computed from Monte Carlo simulations with $J_1=-1, J_2=-0.1$, and $J_3=-0.057$. The temperature $T$ is set to $0.1J_1$ in (c) and (d).}
\end{figure*}

The first key observation is that a range of values exists for the $J_2/J_1$ and $J_3/J_2$ ratios, which is compatible with the experimental measurements.
More specifically, compatibility is found whenever $J_2$ is substantially smaller than $J_1$ ($5J_2 < J_1$), whereas the range of possible values for the $J_3/J_2$ ratio is much more limited ($0.55 < J_3/J_2 < 0.60$, typically) \cite{note_J}.

To test the robustness of these conclusions, we now choose a $\{J_1,J_2,J_3\}=\{-1,-0.1,-0.057\}$ triplet based on the above estimates (see Supplemental Material \cite{SuppMat}) and compare experimental and predicted spin-spin correlators up to the fifteenth neighbors, i.e., up to six lattice parameters (Fig.~\ref{fig4}a).
These correlations are computed as a function of temperature $T$ from ensembles of configurations delivered by Monte Carlo simulations based on single spin flip dynamics.
The results are reported in Figs.~\ref{fig4}b,c in two different manners.
In Fig.~\ref{fig4}b, we show the temperature dependence of the first five spin correlators (colored curves), from the high temperature paramagnetic regime ($T\gg J_1$) in which all magnetic correlations are zero on average, to the lowest temperature we can reach numerically.
Except for the nearest neighbor correlator that will be discussed separately below, the other four experimental correlators (colored dots) fit well with the numerical values at a Monte Carlo temperature $T=0.1J_1$ (Fig.~\ref{fig4}c).
Now fixing the Monte Carlo temperature to $0.1J_1$, spin-spin correlations $C_{2-15}$ obtained numerically (colored bars) agree all extremely well with the experimental values (colored dots).
The agreement is further confirmed by computing the numerical magnetic structure factor at an effective temperature of 0.1 (Fig.~\ref{fig4}d), which agrees well with the experimental one (Fig.~\ref{fig2}a).
In other words, the $\{J_1,J_2,J_3\}$ spin Hamiltonian describes all spin-spin correlations we have measured and indicates that our C$_{60}$ assemblies are at thermodynamic equilibrium, at an effective temperature of $0.1J_1$.
The ground state of this Hamiltonian being a zigzag phase \cite{Mila2016}, the C$_{60}$ assemblies remain in a temperature regime where liquid-like correlations are strong, although a zigzag order starts to develop.

\bigskip

\textit{Magnetic defects and dynamical freezing}--
In the measurements discussed above, the nearest neighbor spin correlator $C_1$ is always found larger than the $-1/3$ value predicted at low temperature.
This means that the local constraint imposed by the antiferromagnetic $J_1$ coupling is not strictly obeyed and local configurations with three molecules ``up'' or three molecules ``down'' are present.
These configurations are magnetic defects.
Analyzing the STM images more closely, we observe that these high-energy configurations are not equally populated: three neighboring molecules are rarely found in the ``up" position, whereas three neighboring molecules are found in a substantial amount in the same ``down" position. 
Zooming in on these three ``down" molecule configurations reveals a distorsion of the triangular lattice (see right panels in Fig~\ref{fig1}b) where the lattice parameter is slightly increased. 
The intermolecular repulsion is not only reduced by a buckling of the C$_{60}$ monolayer, but can also result from a local increase of the in-plane lattice parameter: the C$_{60}$ monolayer has elastic property \cite{Pai2004, note_contrast}.

If this property of the molecular layer is not taken into account in the spin Hamiltonian, it is worth noting that pairwise spin correlations up to the fifteenth neighbor are not affected by the distorsion-induced stabilisation of defects. 
In other words, these defects are rapidly screened and the correlations present in the molecular layer are those of a spin liquid brought to a sufficiently low effective temperature to start developing a zigzag magnetic order.
Although the physics is different, artificial spin ice systems share similar properties. 
For example, despite the presence of a substantial fraction of trapped magnetic defects in artificial square ice magnets, liquid-like spin-spin correlations clearly develop \cite{Perrin2016, Farhan2019, Ostman2018, Brunn2021}.

We emphasize that reaching an effective temperature of the order of $J_2$ (i.e., about 10$\times$ lower than $J_1$) is an intriguing result.
Indeed, the TIAF model is known to require global spin updates to probe low-energy configurations as the single spin flip dynamics freezes \cite{Mila2016}.
Such global updates have no equivalent experimentally and the physical dynamics is likely a single spin flip dynamics.
Our molecular assemblies are thus expected to be found in an arrested configuration at relatively high temperature (in comparison, artificial spin ice systems often freeze when the effective temperature is of the order of the nearest neighbor coupling strength).
What is striking here is that the dynamics freezes at an effective temperature that is not much larger than the lowest effective temperature we reach numerically (see Fig.~\ref{fig4}b).
In fact, if the effective temperature of the C$_{60}$ monolayer was higher, we would have concluded that the disorder phase was close to a pure spin liquid and no trace of the zigzag pattern would have been detected.
Why the thermalization of the C$_{60}$ monolayer is so efficient experimentally remains an open question.

\bigskip

\textit{Conclusion}--
Our work introduces the C$_{60}$/Cu(111) system as a frustrated magnet, despite the absence of magnetism in the two constituent elements.
Surprisingly, although the growth process leading to the formation of the molecular islands is a kinetic process, the corrugated patterns we imaged are reminiscent of a physics at thermodynamic equilibrium, at an effective temperature well below the interaction strength coupling nearest molecules.
This low effective temperature allows us to demonstrate that, counterintuitively, C$_{60}$ molecules interact beyond nearest neighbors.
Our work also demonstrates the relevance of applying a methodology inherited from frustrated magnetism in surface science, especially to detect emerging orders or weak correlations in an apparently disordered and uncorrelated system.
We can envision, for example, to determine to what extent other structural systems \cite{Ganz1991, Carpinelli1998, Ottaviano2003, Azizi2020, Charra1998, Pai2004, Gardener2009, Pasens2015} are well-described by a frustrated $\{J_1, J_2, J_3\}$ spin Hamiltonian, and how their physics compares to the one investigated in arrays of interacting nanostructures \cite{Pip2021}.

\bigskip

\textit{Acknowledgments}--
This work was supported by the Agence Nationale de la Recherche through project no. ANR-17-CE24-0007-03 ``Bio-Ice''. 
M. A.-M. acknowledges the Fondation Nanosciences and the Fondation Universit\'e Grenoble Alpes for support. 
Ana Cristina G\'omez Herrero and Alexandrre Artaud are warmly thank for their help and advices.

\clearpage

\textbf{SUPPLEMENTAL MATERIAL}

\section{Sample growth}

Experiments were performed in an ultra-high vacuum system (base pressure around $1 \times 10^{-10}$ mbar). 
The Cu(111) single crystal was cleaned by repeated sputtering cycles under argon atmosphere (P=$2\times10^{-5}$ mbar, ion energy 1~keV) and annealing (700~K). 
A submonolayer of C$_{60}$ (99.99\% purity) was deposited from a commercial evaporator in a preparation chamber.
The deposition rate was calibrated using a quartz microbalance and different fluxes (from 0.5~\AA/min to 1.7~\AA/min) were used with no significant impact on the growth.
The Cu(111) crystal was kept at room temperature during C$_{60}$ deposition. 
Surface reconstruction of the Cu(111) crystal and lattice parameter of the C$_{60}$ layer were monitored using reflection high-energy electron diffraction.
Room temperature STM imaging was performed in a dedicated chamber, using the constant current mode.
After C$_{60}$ deposition, we always observed the formation of multiple islands (see Fig. \ref{fig1}), sometimes attached to a Cu atomic steps, sometimes in the middle of a terrace. 
The kinetics and details of the growth mechanisms are not taken into account in our analysis. 
What we model is the corrugation disorder within the formed islands.

\section{STM contrast}

The bright/dark STM contrasts are equally populated and no significant unbalance in the height distribution is found in the various C$_{60}$ islands we have imaged.
To quantify the unbalance between dark and bright molecules, we define the ``magnetization" $M$ as the difference between the two molecule states normalized by the total number of molecules. 
For the ten STM images analyzed in this work (Fig. \ref{fig1}), $M=2\% \pm 5\%$.  
Three islands out of ten have a negative magnetization, the others having a positive magnetization.

\begin{figure}
\includegraphics[width=7.9cm]{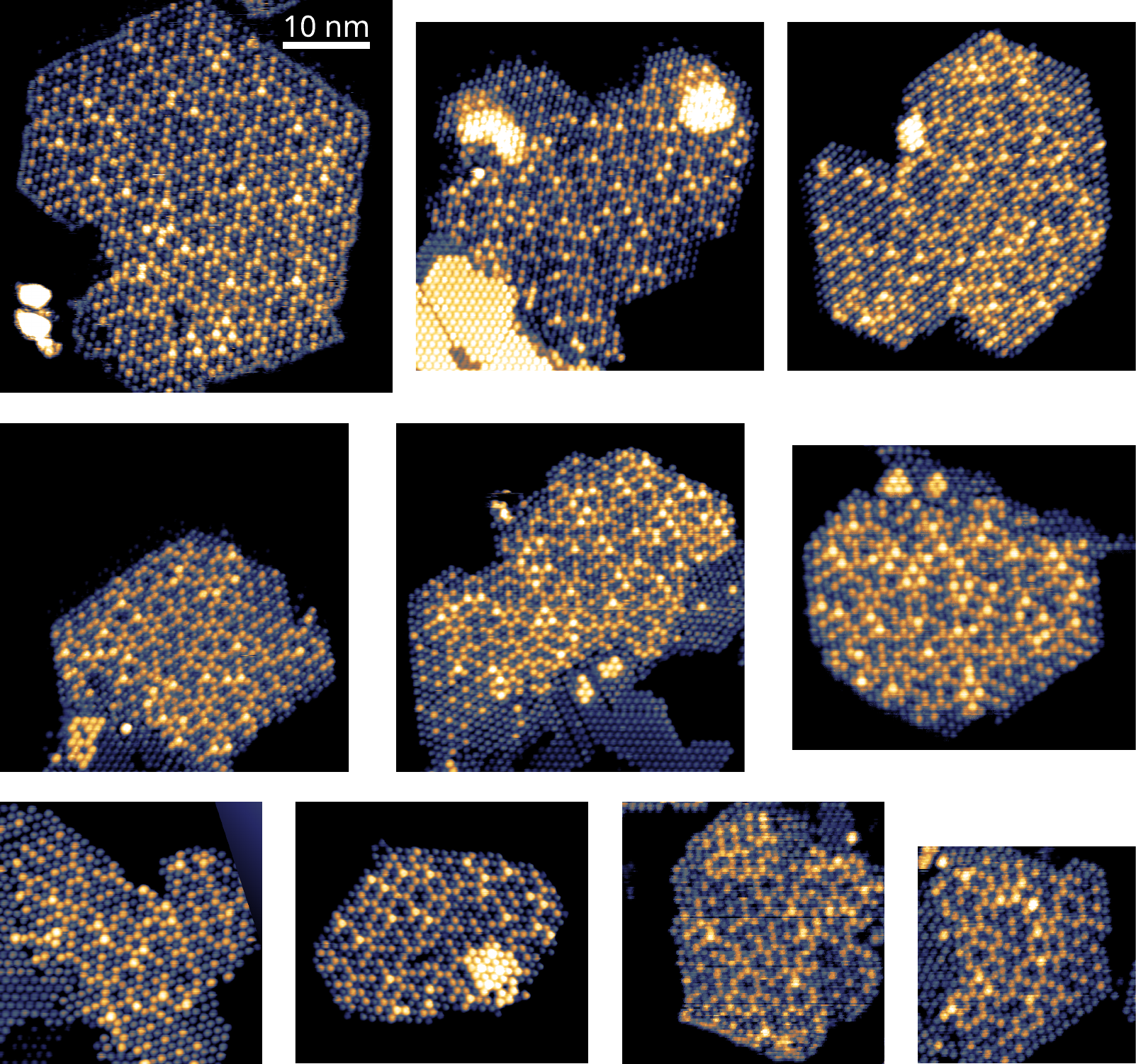}
\caption{\label{fig1} All ten analyzed C$_{60}$ islands exhibiting a disordered corrugation. Molecules appear as bright/dark dots depending on their height. The Cu surface appears in black. Scale bar is 10 nm.}
\end{figure}

\bigskip

\section{Magnetic structure factor}

The magnetic structure factors of the molecular lattices were computed as follow. 
First, we binarized the STM images considering that C$_{60}$ molecules are Ising variables (bright, $\sigma = +1$ or dark, $\sigma = -1$) sitting on a triangular lattice. 
The magnetic structure factor is defined as:
\begin{equation}
	MSF(\vec{q}) = \frac{1}{N}\sum_{ij}\sigma_i \sigma_j e^{i \vec{q} \vec{r}_{ij}}
\end{equation}
where $\vec{q}$ is a vector in reciprocal space, $N$ corresponds to the number of molecules in the lattice and $\vec{r}_{ij} = \vec{r}_j - \vec{r}_j$, with $\vec{r}_i$ and $\vec{r}_j$ the position vectors of molecules sitting on the $i$ and $j$ sites. 
To avoid computing the sum with both $i$ and $j$ indices, the above expression is simplified to obtain the norm of a complex vector:
\begin{equation}
	MSF(\vec{q}) = \frac{1}{N}\lVert \vec{v}_q \rVert ^2 = \frac{1}{N}\vec{v}_q \cdot {\vec{v}_q}*
\end{equation}
with:
\begin{equation}
	\vec{v}_q = \sum_{i}\sigma_i e^{i \vec{q} \vec{r}_i} 
\end{equation}
The magnetic structure factor is computed for each C$_{60}$ island and subsequently averaged over ten STM images to improve statistics (see Fig.~2a in the main text).

\begin{figure}
\includegraphics[width=5cm]{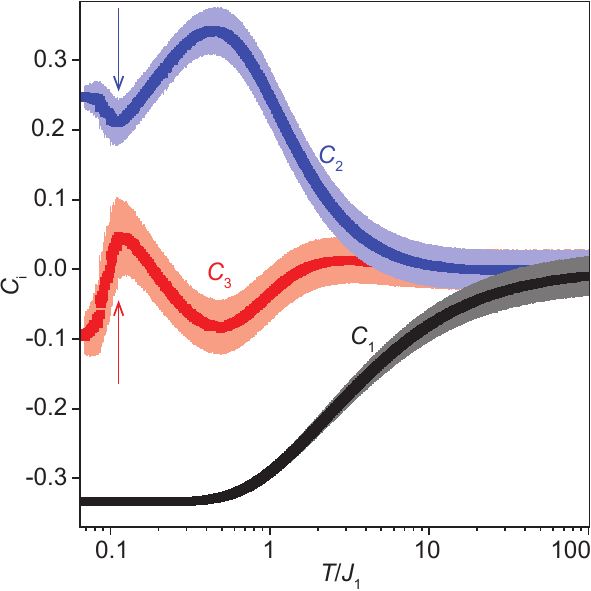}
\caption{\label{fig2} Temperature dependence of the $C_2$ and $C_3$ correlators obtained from Monte Carlo simulations. The local minimum / maximum values we use to fit the experimental measurements are indicated by an arrow.}
\end{figure}

\section{Comparing numerical and experimental values of the $C_2$ and $C_3$ correlators}

To bound the $J_2/J_1$ and $J_3/J_2$ ratios (see Fig.~3 in the main text) we compared the experimental values of the $C_2$ and $C_3$ correlators and the ones obtained numerically \textit{in the low-temperature regime} of the $\{J_1, J_2, J_3\}$ spin Hamiltonian. 
Here we describe how the comparison was made.

As illustrated in Fig.~\ref{fig2}, Monte Carlo simulations reveal that the temperature dependence of the $C_2$ and $C_3$ correlators is not monotonous and varies substantially as the system correlates.
In particular, the $C_2$ / $C_3$ correlators exhibit a local minimum / maximum at low temperature (indicated by an arrow in Fig.~\ref{fig2}).
In our fitting procedure, we spot these local values to match the experimental correlations.
This choice turned out to be efficient as the two local values are very sensitive to a change of the coupling strengths (see the narrow range of the $J_3/J_2$ ratio for which a good match is obtained in Fig.~3b of the main text).
Although this choice might seem arbitrary at first sight, it in fact results from an educated guess based on the observation of the emergent zigzag order in the experimental magnetic structure factor (see Fig.~2a of the main text):

\noindent 1) Such a magnetic order is obtained if $J_3/J_2>0.5$ \textit{and} if the temperature is sufficiently low (see Fig.~\ref{fig3}) \cite{Mila2016}.

\noindent 2) The $C_2$ and $C_3$ correlators are good estimators to distinguish a zigzag pattern from the stripe phase also present in the diagram of the $\{J_1, J_2, J_3\}$ spin Hamiltonian (see Fig.~\ref{fig3}) \cite{Mila2016}.

\noindent We emphasize that the robustness of the fitting procedure was subsequently confirmed by analyzing the first fifteen correlators (see Fig.~4c in the main text).

\begin{figure}
\includegraphics[width=6cm]{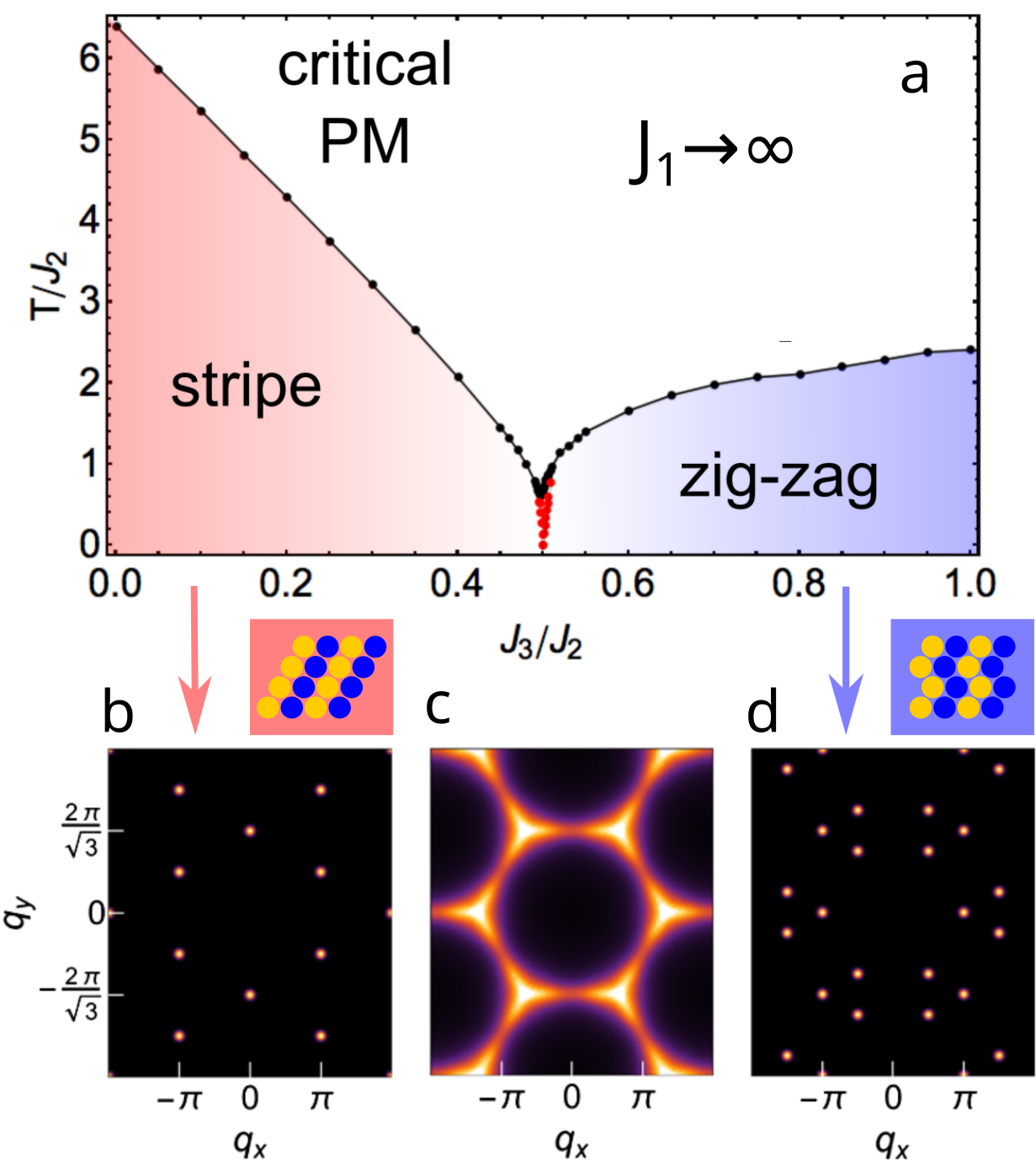}
\caption{\label{fig3} (a) Phase diagram of the $\{J_1\rightarrow \infty, J_2, J_3\}$ TIAF model. (b-d) Magnetic structure factors associated to the stripe (b), critical paramagnet (c) and zigzag (d) phases. Adapted from Ref.\onlinecite{Mila2016}}
\end{figure}

\begin{figure*}
\includegraphics[width=12cm]{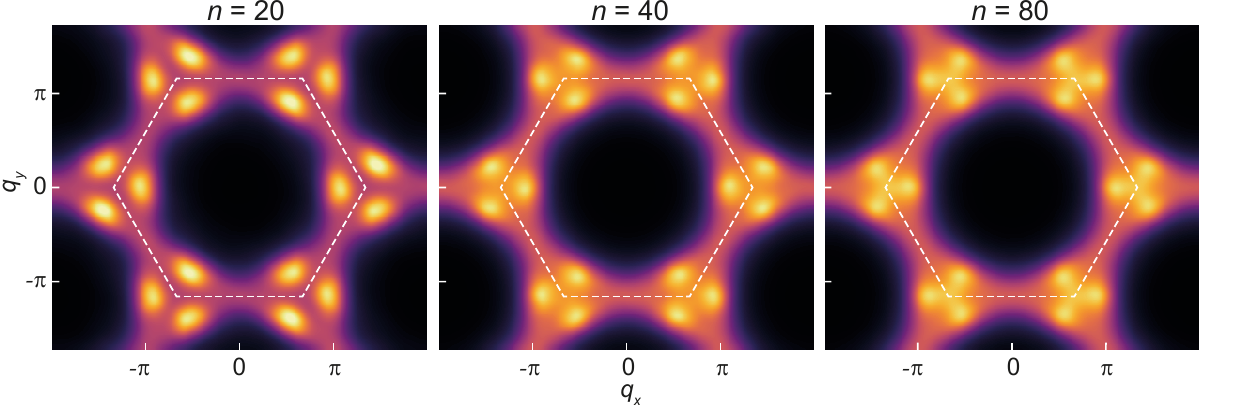}
\caption{\label{fig4} Numerical MSFs computed for $n \times n$ lattices, with $n=$ 20, 40 and 80. The coupling strengths are $\{J_1, J_2, J_3\}=\{-1, -0.1, -0.057\}$ and temperature is $\sim J_2$.}
\end{figure*}

\section{Setting the values of the coupling strengths}

In the main text, we chose the triplet $\{J_1=-1, J_2=-0.1, J_3=-0.057\}$ to fit the experimental data points.
This choice is arbitrary as we can only bound the values of the coupling strengths.
What is critical is to fix the $J_3 / J_2$ ratio in the narrow range of permitted values (see Fig.~3b in the main text).
However, $J_1$ can be changed as long as it remains much larger than $J_2$ (see Fig.~3a in the main text).
For example, choosing the triplet $\{J_1=-100, J_2=-0.1, J_3=-0.057\}$ does not affect our analysis, as also confirmed by the diagram of the $\{J_1\rightarrow \infty, J_2, J_3\}$ TIAF model (see Fig.~\ref{fig3}).

\section{Finite size effects}
In Figs. 2a and 4d of the main manuscript, we note that the emergent Bragg peaks associated to the zigzag phase have elongated shape. 
This is a finite size effect and Bragg peaks become isotropic for larger system sizes.
This is illustrated in Fig.~\ref{fig4} where MSFs are computed for three lattice sizes.

\end{document}